\documentclass[preprint,showpacs,preprintnumbers, 
superscriptaddress,prd]{revtex4}

\usepackage{amsmath,amssymb}
\usepackage{graphicx}
\usepackage{mathrsfs}
\newcommand{\bs}[1]{\ensuremath{\boldsymbol{#1}}}

\begin{document}

\title{Nonchaotic evolution of triangular configuration 
due to gravitational radiation reaction in the three-body problem}
\author{Kei Yamada}
\email{k.yamada@tap.scphys.kyoto-u.ac.jp}
\affiliation{
Department of Physics, Kyoto University, Kyoto 606-8502, Japan} 

\author{Hideki Asada} 
\affiliation{
Faculty of Science and Technology, Hirosaki University,
Hirosaki 036-8561, Japan}

\date{\today}

\begin{abstract}
Continuing work initiated in an earlier publication 
[H. Asada, Phys. Rev. D {\bf 80}, 064021 (2009)], 
the gravitational radiation reaction to 
Lagrange's equilateral triangular solution of 
the three-body problem is investigated in an analytic method.
The previous work is based on the energy balance argument, 
which is sufficient for a two-body system 
because the number of degrees of freedom 
(the semimajor axis and the eccentricity
in quasi-Keplerian cases, for instance) 
equals that of the constants of motion 
such as the total energy and the orbital angular momentum.
In a system with three (or more) bodies,
however, 
the number of degrees of freedom is more than that of the constants of motion.
Therefore, 
the present paper discusses the evolution of the triangular system 
by directly treating the gravitational radiation reaction force to each body.
The perturbed equations of motion are solved 
by using the Laplace transform technique.
It is found that 
the triangular configuration is adiabatically shrinking and 
is kept in equilibrium by
increasing the orbital frequency
due to the radiation reaction 
if the mass ratios satisfy the Newtonian stability condition.
Long-term stability involving 
the first post-Newtonian corrections is also discussed.
\end{abstract}


\maketitle

\section{Introduction}

The first direct detection of gravitational waves, 
named GW150914, 
has been achieved by Advanced LIGO \cite{GWPRL}. 
In the near future, 
gravitational waves astronomy will be largely developed 
by a network of gravitational wave detectors 
such as Advanced VIRGO \cite{aVIRGO} and KAGRA \cite{KAGRA}. 
The test operation, 
named iKAGRA, 
has been started very recently 
as well as Advanced LIGO \cite{aLIGO}.
One of the most promising astrophysical sources is 
inspiraling and merging binary compact stars. 
In fact, 
the GW150914 event fits well with a binary black hole merger \cite{GWPRL}. 
Numerical relativity has succeeded 
in simulating merging neutron stars and black holes 
(e.g. \cite{NR}). 
Analytical methods also have prepared accurate wave form templates 
for inspiraling compact binaries by the post-Newtonian approach \cite{PN}
and also by the black hole perturbations \cite{ST}. 
A lot of effort is placed on 
bridging a gap between the inspiraling stage and the final merging phase 
(e.g., \cite{Damour}).

With growing interest, 
gravitational waves involving three-body interactions 
have been discussed 
(e.g., \cite{CIA, GB, Seto1, DSH}). 
Even the classical three-body (or $N$-body) problem 
in Newtonian gravity admits an increasing number of solutions;
some of them express regular orbits and others are chaotic 
because the number of degrees of freedom of the system is more than 
that of conserved quantities. 
In particular, 
{\it Lagrange's equilateral triangular orbit} 
has stimulated renewed interst 
for relativistic astrophysics 
\cite{THA, Asada, SM, Schnittman, IYA, YA3, YTA, BDEDSG}. 
Very recently, 
a first relativistic hierarchical triple system has been discovered 
by Ransom and his collaborators \cite{Ransom}.  
It has been pointed out by several authors 
that three-body interactions might play 
important roles for compact binary mergers in hierarchical triple systems 
\cite{BLS, MH, Wen, Thompson, Seto2}.

In binary systems, 
the evolution of the semimajor axis and 
the eccentricity is related 
to energy and orbital angular momentum losses 
due to the energy balance argument for the gravitational radiation 
at the second-and-a-half post-Newtonian (2.5PN) order.
Thus, 
one can approximately calculate inspiraling of the binaries 
without directly solving the equation of motion.
In the previous work \cite{Asada}
based on the energy balance argument,
where Lagrange's orbit is assumed to shrink and 
kept in an equilateral triangle, 
Asada has considered the three-body wave forms at the mass quadrupole, 
octupole, and current quadrupole orders, 
especially in an analytic method. 
By using the derived expressions, 
he has solved a gravitational wave inverse problem of 
determining the source parameters to the particular configuration 
(three masses, a distance of the source to an observer, 
and the orbital inclination angle to the line of sight) 
through observations of the gravitational wave forms alone. 
He has discussed also whether and how a binary source can be distinguished 
from a three-body system in Lagrange's orbit or others 
and thus proposed a binary source test.
Strictly speaking, however, 
the energy balance argument is not sufficient 
for three-body systems 
since the number of degrees of freedom in a system with three bodies is 
more than that of the constants of motion.
Hence, one may think that the triangular orbit is likely to
become chaotic owing to the gravitational radiation reaction.
Is the key assumption in the previous work \cite{Asada} correct?

Therefore, 
the main purpose of the present paper is 
to study whether the assumption in the previous work is correct.
Namely, 
the evolution of the orbit is discussed 
through solving directly the equations of motion 
in order to avoid the energy balance argument for Lagrange's orbit. 
In fact, 
even in Newtonian gravity without gravitational radiation, 
it is proved by Gascheau that 
Lagrange's orbit is unstable 
\cite{Gascheau}, 
unless 
\begin{align}
\frac{m_1 m_2 + m_2 m_3 + m_3 m_1}{M^2} < \frac{1}{27} ,
\label{CoSN}
\end{align}
where $m_I ~( I = 1,2,3)$ and $M = \sum m_I$ denote 
the mass of each body and the total mass, respectively. 
This stability condition has recently been corrected 
in the first post-Newtonian (1PN)
approximation as \cite{YTA} 
\begin{align}
\frac{m_1 m_2 + m_2 m_3 + m_3 m_1}{M^2} 
+ \frac{15}{2} \frac{m_1 m_2 m_3}{M^3}  \lambda < 
\frac{1}{27} \left( 1  - \frac{391}{54} \lambda \right) ,
\label{CoS1PN}
\end{align}
where we define 
\begin{align}
\lambda \equiv \left( \frac{G M \omega}{c^3} \right)^{2/3} ,
\end{align}
with 
the common orbital frequency $\omega$ of the system.
Thus, 
the triangular configuration becomes less stable 
by the 1PN corrections. 
In order to 
investigate the effect of the gravitational radiation reaction on 
the evolution of the system, 
first, we focus on the Newtonian stable case that 
the condition \eqref{CoSN} is satisfied. 
Next, we also discuss effects of the 1PN corrections on the stability
by using Eq. \eqref{CoS1PN}.

This paper is organized as follows. 
In Sec. \ref{NaF}, 
we briefly summarize Lagrange's equilateral triangular orbit 
and derive the force by the gravitational radiation reaction. 
In Sec. \ref{Purterbations}, 
we consider the evolution of the orbit due to the radiation reaction.
Section \ref{discussion} is devoted to the discussion.

\section{Notation and Basic Formulation}
\label{NaF}

\subsection{Lagrange's equilateral triangular solution}

First, 
we consider the Newtonian gravity among three bodies 
in coplanar circular motion. 
By using a complex plane for orbital one, 
the location of each body is written as 
\begin{align}
z_I = r_I e^{i \theta_I} ,
\end{align}
where $r_I$ and $\theta_I$ are 
the field point distance and direction of the $I$th body.
We choose the origin of the coordinates as the center of mass 
so that 
\begin{align}
m_1 z_1 + m_2 z_2 + m_3 z_3 = 0 ,
\end{align}
and we denote the relative position between bodies 
$z_{I J} \equiv z_I - z_J$ as 
\begin{align}
z_{I J} = r_{I J} e^{i \theta_{I J}} ,
\end{align}
where $r_{I J}$ and $\theta_{I J}$ are 
the separation and the relative angle between the bodies, respectively.
Hence, 
we have,  
with $I \neq J, J \neq K, K \neq I$,
\begin{align}
r_I e^{i \theta_I} = 
\nu_J r_{I J} e^{i \theta_{I J}} 
+ \nu_K r_{I K} e^{i \theta_{I K}} ,
\label{I-to-IJ}
\end{align}
where $\nu_I \equiv m_I /M$ is the mass ratio.

We consider an equilateral triangular configuration in equilibrium. 
Namely, we denote $r_{I J} = \ell = {\rm constant}$ and 
\begin{align}
\theta_{23} = \theta_{12} + \frac23 \pi , 
\label{t23-t12} \\
\theta_{31} = \theta_{12} - \frac23 \pi . 
\label{t31-t12}
\end{align}
Then, the equation of motion for $z_{I J}$ becomes
\begin{align}
\left( \frac{d \theta_{I J}}{d t} \right)^2 
+ i \frac{d^2 \theta_{I J}}{d t^2} 
&= \frac{G M}{\ell^3} . 
\end{align} 
Thus, 
each body can move around the center of mass with the orbital frequency 
\begin{align}
\omega \equiv \frac{d \theta_{I J}}{d t} 
= \sqrt{\frac{G M}{\ell^3}} .
\end{align}
By using Eqs. \eqref{I-to-IJ}--\eqref{t31-t12}, 
we obtain
\begin{align}
\bar{r}_I \equiv \frac{r_I}{\ell} 
= \sqrt{\nu_J^2 + \nu_J \nu_K + \nu_K^2} .
\end{align}

In this paper, 
in order to investigate the effect of the gravitational radiation reaction on 
the evolution of the system, 
we focus on the case that the Newtonian equilateral triangle is stable, 
namely, 
\begin{align}
V \equiv \nu_1 \nu_2 + \nu_2 \nu_3 + \nu_3 \nu_1 < \frac{1}{27} .
\label{cond.Newton}
\end{align}

\subsection{Gravitational radiation reaction to Lagrange's orbit}

The reaction force due to the gravitational quadrupole radiation 
to the $I$th body per unit mass in the harmonic gauge is expressed as 
(see Appendix \ref{app1})
\begin{align}
F^{\rm RR}_I = - \frac{32}{5} \frac{G M}{\ell^2} \varepsilon \bar{r}_I  
[ A_I + i B_I ] e^{i \theta_I} ,
\label{RR-Force}
\end{align}
where we define 
\begin{align}
\varepsilon \equiv 
\left( \frac{G M \omega}{c^3} \right)^{5/3} ,
\end{align}
and 
\begin{align}
A_I &\equiv \sum_J \nu_J ( \bar{r}_J )^2 
\sin (2 \theta_I - 2 \theta_J) , 
\label{A} \\
B_I &\equiv \sum_J \nu_J ( \bar{r}_J )^2 
\cos (2 \theta_I - 2 \theta_J) .
\label{B}
\end{align}
In the case of the equilateral triangular configuration, 
$\bar{r}_J$ and $\theta_I - \theta_J$ are constant, 
and hence, 
$A_I$ and $B_I$ in Eq. \eqref{RR-Force} are constant 
at the first order of $\varepsilon$.
Moreover, one can show
\begin{align}
\sum_I F^{\rm RR}_I = 0 .
\end{align}
It follows that 
the position of the center of mass is not changed by the reaction force.
Using Eq. \eqref{I-to-IJ}, 
the reaction force to $z_{I J}$ is expressed as 
\begin{align}
F^{\rm RR}_{I J} &\equiv F^{\rm RR}_I - F^{\rm RR}_J 
\notag\\
&= \frac{16}{5} \frac{G M}{\ell^2} \varepsilon 
( A_{I J} - i B_{I J} ) e^{i \theta_{I J}} ,
\end{align}
where 
\begin{align}
A_{I J} &= \sqrt{3} (\nu_I - \nu_J ) \nu_K , 
\label{ex-AIJ} \\
B_{I J} &= \nu_I ( \nu_J - \nu_K ) + \nu_J ( \nu_K - \nu_I ) .
\label{ex-BIJ}
\end{align}

\section{Evolution of Lagrange's orbit}
\label{Purterbations}

\begin{figure}[h]
\begin{center}
  \includegraphics[width=10cm]{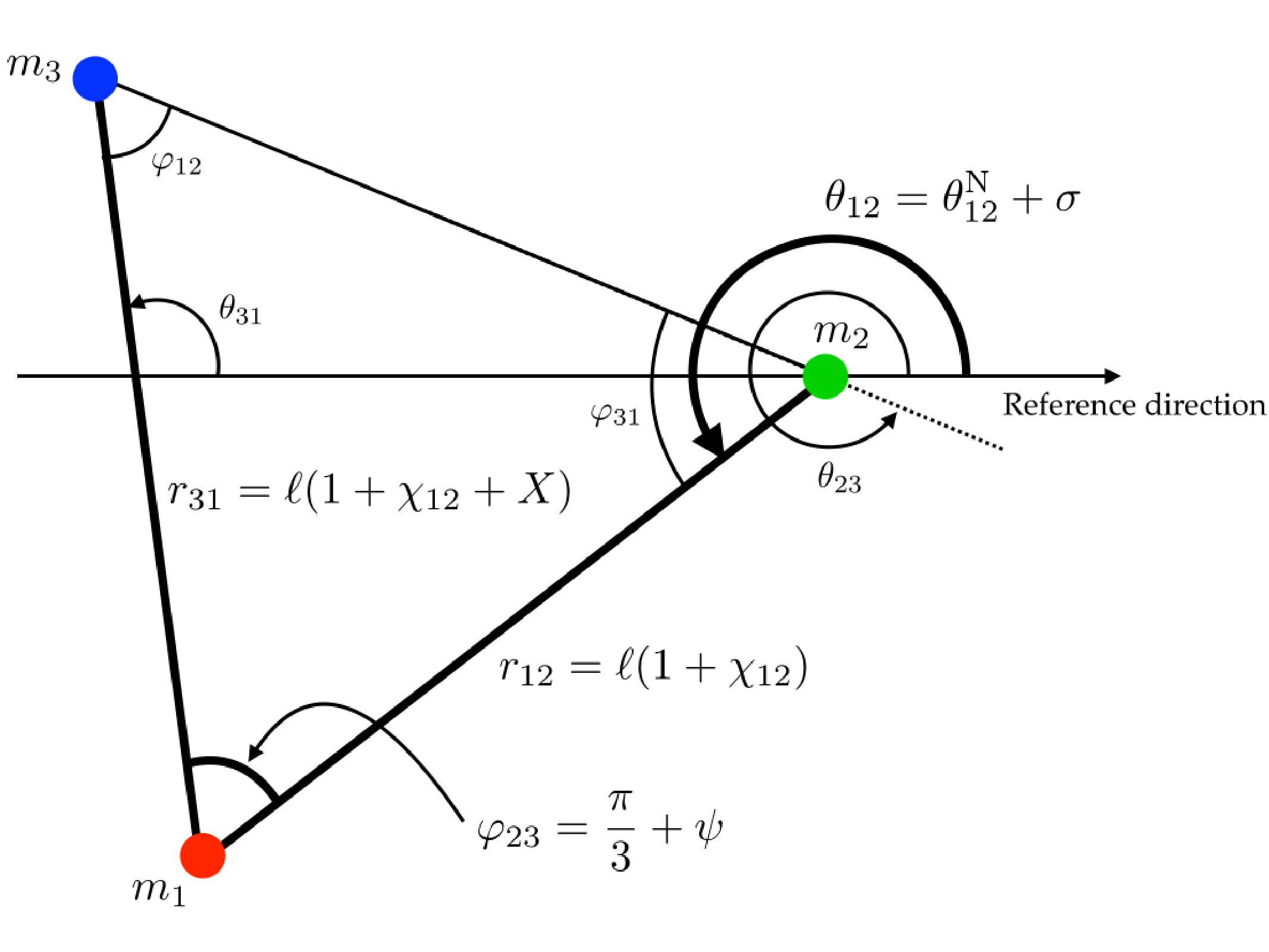}
\caption{
Perturbations in Lagrange's orbit.
Hence, $\chi_{12}$ and $\sigma$ correspond to 
the scale transformation of the triangle 
and the change of the angle of the system to a reference direction, 
respectively.
On the other hand, 
$X$ and $\psi$ are
the degrees of freedom of a shape change
from the equilateral triangle.}
\label{fig-1}
\end{center}
\end{figure}

The motion of each body and that of 
the relative positions are perturbed due to gravitational radiation, 
while the position of the common center of mass and 
the orbital plane does not change.
Therefore, the number of degrees of freedom 
for the perturbations in  Lagrange's orbit is 4.
Let us consider the perturbation variables ($\chi_{12}, X, \psi, \sigma$), 
so that 
\begin{align}
r_{12} = \ell &\to 
\ell ( 1 + \chi_{12} ) , \\
r_{31} = \ell &\to 
\ell ( 1 + \chi_{12} + X ), \\
\varphi_{23} = \frac{\pi}{3} &\to
\frac{\pi}{3} + \psi , \\
\theta_{12} = \theta^{\rm N}_{12} &\to 
\theta^{\rm N}_{12} + \sigma ,
\end{align}
where $\varphi_{23}$ denotes the opposite angle to $r_{23}$ 
and $\theta^{\rm N}_{12}$ is the Newtonian value (Fig. \ref{fig-1}). 
In this choice of variables, 
$\chi_{12}$ and $\sigma$ correspond to 
the scale transformation of the triangle 
and the change of the angle of the system to a reference direction, 
respectively.
On the other hand, 
$X$ and $\psi$ are
the degrees of freedom of a shape change from the equilateral triangle. 
Therefore, 
the shrinking triangular configuration will 
adiabatically stay 
in equilibrium 
if and only if both $X$ and $\psi$ do not increase with time.
We suppose that 
the order of magnitude of all the perturbations is $\varepsilon$.

The perturbed equations of motion are expressed as 
\begin{align}
\ddot{\chi}_{12} - 3 \chi_{12} - 2 \dot{\sigma} 
- \frac94 \nu_3 X - \frac{3 \sqrt{3}}{4} \nu_3 \psi 
- \frac{16}{5} \varepsilon A_{12} &= 0 , 
\label{EoM12Re} \\ 
2 \dot{\chi}_{12} + \ddot{\sigma} 
- \frac{3 \sqrt{3}}{4} \nu_3 X + \frac94 \nu_3 \psi 
+ \frac{16}{5} \varepsilon B_{12} &= 0 , 
\label{EoM12Im} \\
\ddot{\chi}_{12} - 3 \chi_{12} - 2 \dot{\sigma} 
+ \ddot{X} - \left( 3 - \frac94 \nu_2 \right) X 
- 2 \dot{\psi} - \frac{3 \sqrt{3}}{4} \nu_2 \psi 
- \frac{16}{5} \varepsilon A_{31} &= 0 , 
\label{EoM31Re} \\ 
2 \dot{\chi}_{12} + \ddot{\sigma} 
+ 2 \dot{X} - \frac{3 \sqrt{3}}{4} \nu_2 X 
+ \ddot{\psi} - \frac94 \nu_2 \psi 
+ \frac{16}{5} \varepsilon B_{31} &= 0 ,
\label{EoM31Im}
\end{align}
where the dot denotes the derivative 
with respect to a normalized time $\bar{t} \equiv \omega t$.
These equations do not contain $\sigma$. 
This is consistent with the fact that 
the initial value of $\sigma$ can be zero 
through the appropriate coordinate rotation.
In order to avoid such a redundancy, 
let us use the perturbation in the orbital frequency $\omega$ as
\begin{align}
\varpi \equiv \dot{\sigma} ,
\end{align}
instead of $\sigma$ as usual. 

By solving Eqs. \eqref{EoM12Re}--\eqref{EoM31Im} 
(see Appendix \ref{solve} for more detail), 
we obtain 
\begin{align}
X &= 
\frac{16}{45 V} \varepsilon 
\left( 
3 (\nu_2 + \nu_3) (A_{12} - A_{31}) + \sqrt{3} (\nu_2 - \nu_3) (B_{12} - B_{31}) 
\right) 
+ X_{\rm (osc.)} , 
\label{ex-X} \\
\psi &= 
- \frac{16}{45 V} \varepsilon 
\left( 
\sqrt{3} (\nu_2 - \nu_3) (A_{12} - A_{31}) 
+ (4 - 3 \nu_2 - 3 \nu_3) (B_{12} - B_{31}) 
\right) 
+ \psi_{\rm (osc.)} ,
\label{ex-psi} \\
\chi_{12} &= 
\frac{16}{5} \varepsilon \bar{t} 
\left[
- 2 V B_{12} 
+ \sqrt{3} \nu_2 \nu_3 (A_{12} - A_{31}) 
+ \nu_3 ( 2 - \nu_2 - 2 \nu_3 ) (B_{12} - B_{31})
\right]
\notag\\
&~~~
+ \frac{8}{45 V} \varepsilon 
\left[
18 V A_{12} - 3 \nu_3 ( 8 - 3 \nu_2 - 6 \nu_3 ) (A_{12} - A_{31}) 
+ \sqrt{3} \nu_3 (2 + 9 \nu_2 ) (B_{12} - B_{31})
\right]
\notag\\
&~~~
+ 4 \chi_{12 \, {\rm (ini.)}} + 2 \varpi_{\rm (ini.)} 
+ \frac{2 ( 2 - \nu_2 - 2 \nu_3 ) \nu_3}{V} X_{\rm (ini.)} 
- \frac{\sqrt{3} \nu_2 \nu_3}{V} \dot{X}_{\rm (ini.)}
\notag\\
&~~~
+ \frac{2 \sqrt{3} \nu_2 \nu_3}{V} \psi_{\rm (ini.)}
+ \frac{( 2 - \nu_2 - 2 \nu_3) \nu_3}{V} \dot{\psi}_{\rm (ini.)}
+ \chi_{12 \, {\rm (osc.)}} , 
\label{ex-chi12} \\
\varpi &=
- \frac{24}{5 V} \varepsilon \bar{t} \left[
- 2 V B_{12} + \sqrt{3} \nu_2 \nu_3 (A_{12} - A_{31}) 
+ \nu_3 ( 2 - \nu_2 - 2 \nu_3 ) (B_{12} - B_{31}) 
\right]
\notag\\
&~~~
+ \frac{16}{5 V} \varepsilon 
\left( 
- 2 V A_{12} + \nu_3 ( 2 - \nu_2 - 2 \nu_3 ) (A_{12} - A_{31}) 
- \sqrt{3} \nu_2 \nu_3 (B_{12} - B_{31}) \right)
\notag\\
&~~~
- \frac{3}{2} \left( 4 \chi_{12 \, {\rm (ini.)}} + 2 \varpi_{\rm (ini.)} 
+ \frac{2 \nu_3 (2 - \nu_2 - 2 \nu_3)}{V} X_{\rm (ini.)} 
- \frac{\sqrt{3} \nu_2 \nu_3}{V} \dot{X}_{\rm (ini.)} 
\right.
\notag\\
&~~~
\left. 
+ \frac{2 \sqrt{3} \nu_2 \nu_3}{V} \psi_{\rm (ini.)} 
+ \frac{\nu_3 (2 - \nu_2 - 2 \nu_3)}{V} \dot{\psi}_{\rm (ini.)} \right)
+ \varpi_{\rm (osc.)}
\label{ex-varpi} ,
\end{align}
where the subscript (ini.) denotes the initial value 
and $X_{\rm (osc.)}$, $\psi_{\rm (osc.)}$, 
$\chi_{12 \, {\rm (osc.)}}$, and $\varpi_{\rm (osc.)}$ are oscillating terms 
expressed as 
\begin{align}
X_{\rm (osc.)} &= 
\frac{1}{(\alpha - \beta) (\alpha + \beta)} 
\left[
\left( \alpha^2 - 4 + \frac94 (\nu_2 + \nu_3) \right) X_{\rm (ini.)} 
- \frac{3 \sqrt{3}}{4} (\nu_2 - \nu_3) \psi_{\rm (ini.)} - 2 \dot{\psi}_{\rm (ini.)} 
\right.
\notag\\
&~~~
\left.
+ \frac{16}{5 \alpha^2} \varepsilon 
\left( \left[ \alpha^2 + \frac94 (\nu_2 + \nu_3) \right] (A_{12} - A_{31}) 
+ \frac{3 \sqrt{3}}{4} (\nu_2 - \nu_3) (B_{12} - B_{31}) \right)
\right] \cos (\alpha \bar{t})
\notag\\
&~~~
- \frac{1}{(\alpha - \beta) (\alpha + \beta)} 
\left[
\left( \beta^2 - 4 + \frac94 (\nu_2 + \nu_3) \right) X_{\rm (ini.)} 
- \frac{3 \sqrt{3}}{4} (\nu_2 - \nu_3) \psi_{\rm (ini.)} - 2 \dot{\psi}_{\rm (ini.)} 
\right.
\notag\\
&~~~
\left.
+ \frac{16}{5 \beta^2} \varepsilon 
\left( \left[ \beta^2 + \frac94 (\nu_2 + \nu_3) \right] (A_{12} - A_{31}) 
+ \frac{3 \sqrt{3}}{4} (\nu_2 - \nu_3) (B_{12} - B_{31}) \right)
\right] \cos (\beta \bar{t}) 
\notag\\
&~~~
- \frac{1}{\alpha (\alpha - \beta) (\alpha + \beta)} 
\left[
\frac{3 \sqrt{3}}{4} (\nu_2 - \nu_3) ( 2 X_{\rm (ini.)} + \dot{\psi}_{\rm (ini.)} ) 
- \alpha^2 \dot{X}_{\rm (ini.)} 
\right.
\notag\\
&~~~
\left.
- \frac94 (\nu_2 + \nu_3) (\dot{X}_{\rm (ini.)} - 2 \psi_{\rm (ini.)})
+ \frac{32}{5} \varepsilon (B_{12} - B_{31})
\right] \sin (\alpha \bar{t})
\notag\\
&~~~
+ \frac{1}{\beta (\alpha - \beta) (\alpha + \beta)} 
\left[
\frac{3 \sqrt{3}}{4} (\nu_2 - \nu_3) ( 2 X_{\rm (ini.)} + \dot{\psi}_{\rm (ini.)} ) 
- \beta^2 \dot{X}_{\rm (ini.)} 
\right.
\notag\\
&~~~
\left.
- \frac94 (\nu_2 + \nu_3) (\dot{X}_{\rm (ini.)} - 2 \psi_{\rm (ini.)})
+ \frac{32}{5} \varepsilon (B_{12} - B_{31})
\right] \sin (\beta \bar{t}) , \\
\psi_{\rm (osc.)} &= 
- \frac{1}{(\alpha - \beta) (\alpha + \beta)} \left[
\frac{3 \sqrt{3}}{4} (\nu_2 - \nu_3) X_{\rm (ini.)} - 2 \dot{X}_{\rm (ini.)} 
- \left( \alpha^2 - 1 - \frac94 (\nu_2 + \nu_3) \right) \psi_{\rm (ini.)}
\right.
\notag\\
&~~~
+ \left.
\frac{16}{5 \alpha^2} \varepsilon 
\left( \frac{3 \sqrt{3}}{4} (\nu_2 - \nu_3) (A_{12} - A_{31})
+ \left[ \alpha^2 + 3 - \frac94 (\nu_2 + \nu_3) \right] (B_{12} - B_{31})
\right)
\right] \cos (\alpha \bar{t}) 
\notag\\
&~~~
+ \frac{1}{(\alpha - \beta) (\alpha + \beta)} \left[
\frac{3 \sqrt{3}}{4} (\nu_2 - \nu_3) X_{\rm (ini.)} - 2 \dot{X}_{\rm (ini.)} 
- \left( \beta^2 - 1 - \frac94 (\nu_2 + \nu_3) \right) \psi_{\rm (ini.)}
\right.
\notag\\
&~~~
+ \left.
\frac{16}{5 \beta^2} \varepsilon 
\left( \frac{3 \sqrt{3}}{4} (\nu_2 - \nu_3) (A_{12} - A_{31})
+ \left[ \beta^2 + 3 - \frac94 (\nu_2 + \nu_3) \right] (B_{12} - B_{31})
\right)
\right] \cos (\beta \bar{t}) 
\notag\\
&~~~
+ \frac{1}{\alpha (\alpha - \beta) (\alpha + \beta)} \left[
\left( 3 - \frac94 (\nu_2 + \nu_3) \right) 
(2 X_{\rm (ini.)} + \dot{\psi}_{\rm (ini.)}) 
\right.
\notag\\
&~~~
\left.
- \frac{3 \sqrt{3}}{4} (\nu_2 - \nu_3) (\dot{X}_{\rm (ini.)} - 2 \psi_{\rm (ini.)}) 
+ \alpha^2 \dot{\psi}_{\rm (ini.)} 
- \frac{32}{5} \varepsilon (A_{12} - A_{31})
\right] \sin (\alpha \bar{t}) 
\notag\\
&~~~
- \frac{1}{\beta (\alpha - \beta) (\alpha + \beta)} \left[
\left( 3 - \frac94 (\nu_2 + \nu_3) \right) 
(2 X_{\rm (ini.)} + \dot{\psi}_{\rm (ini.)}) 
\right.
\notag\\
&~~~
\left.
- \frac{3 \sqrt{3}}{4} (\nu_2 - \nu_3) (\dot{X}_{\rm (ini.)} - 2 \psi_{\rm (ini.)}) 
+ \beta^2 \dot{\psi}_{\rm (ini.)} 
- \frac{32}{5} \varepsilon (A_{12} - A_{31})
\right] \sin (\beta \bar{t}) ,
\end{align}

\begin{align}
\chi_{12 \, {\rm (osc.)}} &= 
- \left[
3 \chi_{12 \, {\rm (ini.)}} + 2 \varpi_{\rm (ini.)} 
- \frac{1}{2 V} \left( - \nu_3 ( 2 \nu_1 + \nu_2 ) 
( 3 X_{\rm (ini.)} + 2 \dot{\psi}_{\rm (ini.)} )
+ \sqrt{3} \nu_2 \nu_3 ( 2 \dot{X}_{\rm (ini.)} - 3 \psi_{\rm (ini.)} ) \right) 
\right.
\notag\\
&~~~
\left.
+ \frac{8}{5 V} \varepsilon 
\left( 
2 V A_{12} -  \nu_3 ( 2 \nu_1 + \nu_2 ) (A_{12} - A_{31}) 
+ \sqrt{3} \nu_2 \nu_3 (B_{12} - B_{31}) \right)
\right] \cos \bar{t} 
\notag\\
&~~~
+ \left[
\dot{\chi}_{12 \, {\rm (ini.)}} 
+ \frac{1}{2 V} \left( \nu_3 ( 2 \nu_1 + \nu_2 ) \dot{X}_{\rm (ini.)} 
+ \sqrt{3} \nu_2 \nu_3 \dot{\psi}_{\rm (ini.)} \right)
\right.
\notag\\
&~~~
\left.
+ \frac{16}{5 V} \varepsilon 
\left( 2 V B_{12} - \sqrt{3} \nu_2 \nu_3 (A_{12} - A_{31}) 
- \nu_3 ( 2 \nu_1 + \nu_2 ) (B_{12} - B_{31}) \right) 
\right] \sin \bar{t}
\notag\\
&~~~
+ \frac{\sqrt{3}}{18 ( \alpha - \beta ) (\alpha + \beta ) V} 
\notag\\
&~~~
\times 
\left[
\sqrt{3} \alpha^2 \nu_3 ( 2 \alpha^2 - 8 + 3 \nu_2 + 6 \nu_3 ) X_{\rm (ini.)} 
+ 6 \sqrt{3} \nu_3 ( 2 - \nu_2 - 2 \nu_3 ) 
( 2 X_{\rm (ini.)} + \dot{\psi}_{\rm (ini.)} ) 
\right.
\notag\\
&~~~
- 18 \nu_2 \nu_3 ( \dot{X}_{\rm (ini.)} - 2 \psi_{\rm (ini.)} ) 
+ \alpha^2 \nu_3 ( 2 \alpha^2 - 2 - 9 \nu_2 ) \psi_{\rm (ini.)}
\notag\\
&~~~
+ \left.
\frac{16}{5} \varepsilon 
\left( \sqrt{3} \nu_3 ( 2 \alpha^2 - 8 + 3 \nu_2 + 6 \nu_3 ) (A_{12} - A_{31}) 
- \nu_3 ( 2 \alpha^2 - 2 - 9 \nu_2 ) (B_{12} - B_{31}) \right)
\right] \cos (\alpha \bar{t})
\notag\\
&~~~
- \frac{\sqrt{3}}{18 ( \alpha - \beta ) (\alpha + \beta ) V} 
\notag\\
&~~~
\times 
\left[
\sqrt{3} \beta^2 \nu_3 ( 2 \beta^2 - 8 + 3 \nu_2 + 6 \nu_3 ) X_{\rm (ini.)} 
+ 6 \sqrt{3} \nu_3 ( 2 \nu_1 + \nu_2 ) 
( 2 X_{\rm (ini.)} + \dot{\psi}_{\rm (ini.)} ) 
\right.
\notag\\
&~~~
- 18 \nu_2 \nu_3 ( \dot{X}_{\rm (ini.)} - 2 \psi_{\rm (ini.)} ) 
+ \beta^2 \nu_3 ( 2 \beta^2 - 2 - 9 \nu_2 ) \psi_{\rm (ini.)}
\notag\\
&~~~
+ \left.
\frac{16}{5} \varepsilon 
\left( \sqrt{3} \nu_3 ( 2 \beta^2 - 8 + 3 \nu_2 + 6 \nu_3 ) (A_{12} - A_{31}) 
- \nu_3 ( 2 \beta^2 - 2 - 9 \nu_2 ) (B_{12} - B_{31}) \right)
\right] \cos (\beta \bar{t})
\notag\\
&~~~
- \frac{\sqrt{3}}{18 \alpha ( \alpha - \beta ) (\alpha + \beta ) V} 
\notag\\
&~~~
\times 
\left[
\frac{27 V}{2} \nu_3 \left( ( 2 X_{\rm (ini.)} + \dot{\psi}_{\rm (ini.)} ) 
+ \sqrt{3} ( \dot{X}_{\rm (ini.)} - 2 \psi_{\rm (ini.)} ) \right) 
\right.
\notag\\
&~~~
\left.
+ 3 \sqrt{3} \alpha^2 \nu_3 ( 2 - \nu_2 - 2 \nu_3 ) \dot{X}_{\rm (ini.)} 
+ 9 \alpha^2 \nu_2 \nu_3 \dot{\psi}_{\rm (ini.)}
\right.
\notag\\
&~~~
\left.
- \frac{96 \sqrt{3}}{5} \nu_3 \varepsilon 
\left( \sqrt{3} \nu_2 (A_{12} - A_{31}) 
+ ( 2 - \nu_2 - 2 \nu_3 ) (B_{12} - B_{31}) \right)
\right] \sin (\alpha \bar{t})
\notag\\
&~~~
+ \frac{\sqrt{3}}{18 \beta ( \alpha - \beta ) (\alpha + \beta ) V} 
\notag\\
&~~~
\times 
\left[
\frac{27 V}{2} \nu_3 \left( ( 2 X_{\rm (ini.)} + \dot{\psi}_{\rm (ini.)} ) 
+ \sqrt{3} ( \dot{X}_{\rm (ini.)} - 2 \psi_{\rm (ini.)} ) \right) 
\right.
\notag\\
&~~~
\left.
+ 3 \sqrt{3} \beta^2 \nu_3 ( 2 \nu_1 + \nu_2 ) \dot{X}_{\rm (ini.)} 
+ 9 \beta^2 \nu_2 \nu_3 \dot{\psi}_{\rm (ini.)}
\right.
\notag\\
&~~~
\left.
- \frac{96 \sqrt{3}}{5} \nu_3 \varepsilon 
\left( \sqrt{3} \nu_2 (A_{12} - A_{31}) 
+ ( 2 - \nu_2 - 2 \nu_3 ) (B_{12} - B_{31}) \right)
\right] \sin (\beta \bar{t}) , 
\end{align}

\begin{align}
\varpi_{\rm (osc.)} &= 
2 \left[
3 \chi_{12 \, {\rm (ini.)}} + 2 \varpi_{\rm (ini.)} 
- \frac{1}{2 V} \left( - \nu_3 ( 2 \nu_1 + \nu_2 ) 
( 3 X_{\rm (ini.)} + 2 \dot{\psi}_{\rm (ini.)} )
+ \sqrt{3} \nu_2 \nu_3 ( 2 \dot{X}_{\rm (ini.)} - 3 \psi_{\rm (ini.)} ) \right) 
\right.
\notag\\
&~~~
\left.
+ \frac{8}{5 V} \varepsilon 
\left( 
2 V A_{12} -  \nu_3 ( 2 \nu_1 + \nu_2 ) (A_{12} - A_{31}) 
+ \sqrt{3} \nu_2 \nu_3 (B_{12} - B_{31}) \right)
\right] \cos \bar{t} 
\notag\\
&~~~
- 2 \left[
\dot{\chi}_{12 \, {\rm (ini.)}} 
+ \frac{1}{2 V} \left( \nu_3 ( 2 \nu_1 + \nu_2 ) \dot{X}_{\rm (ini.)} 
+ \sqrt{3} \nu_2 \nu_3 \dot{\psi}_{\rm (ini.)} \right)
\right.
\notag\\
&~~~
\left.
+ \frac{16}{5 V} \varepsilon 
\left( 2 V B_{12} - \sqrt{3} \nu_2 \nu_3 (A_{12} - A_{31}) 
- \nu_3 ( 2 \nu_1 + \nu_2 ) (B_{12} - B_{31}) \right) 
\right] \sin \bar{t} 
\notag\\
&~~~
+ \frac{\sqrt{3}}{18 ( \alpha - \beta ) (\alpha + \beta ) V} 
\left[
- \sqrt{3} \nu_3 
\left[ 2 \alpha^4 - 2 \alpha^2 + 27 ( 2 \nu_1 + \nu_2 ) \right] 
( 2 X_{\rm (ini.)} + \dot{\psi}_{\rm (ini.)} )
+ 9 \alpha^2 \nu_2 \nu_3 \dot{X}_{\rm (ini.)} 
\right.
\notag\\
&~~~
+ \nu_3 ( 2 \alpha^4 - 2 \alpha^2 + 27 \nu_2 ) 
( \dot{X}_{\rm (ini.)} - 2 \psi_{\rm (ini.)} ) 
- 3 \sqrt{3} \alpha^2 \nu_3 ( 2 \nu_1 + \nu_2 ) \dot{\psi}_{\rm (ini.)} 
\notag\\
&~~~
\left.
+ \frac{96 \sqrt{3}}{5} \nu_3 \varepsilon 
\left( ( 2 \nu_1 + \nu_2 ) (A_{12} - A_{31}) 
- \sqrt{3} \nu_2 (B_{12} - B_{31})
\right)
\right] \cos (\alpha \bar{t})
\notag\\
&~~~
- \frac{\sqrt{3}}{18 ( \alpha - \beta ) (\alpha + \beta ) V} 
\left[
- \sqrt{3} \nu_3 
\left[ 2 \beta^4 - 2 \beta^2 + 27 ( 2 \nu_1 + \nu_2 ) \right] 
( 2 X_{\rm (ini.)} + \dot{\psi}_{\rm (ini.)} )
+ 9 \beta^2 \nu_2 \nu_3 \dot{X}_{\rm (ini.)} 
\right.
\notag\\
&~~~
+ \nu_3 ( 2 \beta^4 - 2 \beta^2 + 27 \nu_2 ) 
( \dot{X}_{\rm (ini.)} - 2 \psi_{\rm (ini.)} ) 
- 3 \sqrt{3} \beta^2 \nu_3 ( 2 \nu_1 + \nu_2 ) \dot{\psi}_{\rm (ini.)} 
\notag\\
&~~~
\left.
+ \frac{96 \sqrt{3}}{5} \nu_3 \varepsilon 
\left( ( 2 - \nu_2 - 2 \nu_3 ) (A_{12} - A_{31}) 
- \sqrt{3} \nu_2 (B_{12} - B_{31})
\right)
\right] \cos (\beta \bar{t})
\notag\\
&~~~
+ \frac{\sqrt{3}}{18 \alpha ( \alpha - \beta ) (\alpha + \beta ) V} \left[
\frac{27}{4} V \nu_3 \left( ( 2 \alpha^2 + 9 \nu_2 ) X_{\rm (ini.)} 
- \sqrt{3} ( 2 \alpha^2 + 6 - 3 \nu_2 - 6 \nu_3 ) \psi_{\rm (ini.)} \right) 
\right.
\notag\\
&~~~+ 6 \sqrt{3} \alpha^2 \nu_3 ( 2 - \nu_2 - 2 \nu_3 ) \dot{X}_{\rm (ini.)} 
+ 18 \alpha^2 \nu_2 \nu_3 \dot{\psi}_{\rm (ini.)}
\notag\\
&~~~
- \frac{16}{5} \nu_3 \varepsilon 
\left( \left[ 2 \alpha^4 - ( 2 - 9 \nu_2 ) \alpha^2 + 27 \nu_2 \right] 
(A_{12} - A_{31}) 
\right.
\notag\\
&~~~
\left.
\left.
+ \sqrt{3} \left[ 2 \alpha^4 + ( 4 - 3 \nu_2 - 6 \nu_3 ) \alpha^2 
+ 9 ( 2 \nu_1 + \nu_2 ) \right] (B_{12} - B_{31}) \right)
\right] \sin (\alpha \bar{t})
\notag\\
&~~~
- \frac{\sqrt{3}}{18 \beta ( \alpha - \beta ) (\alpha + \beta ) V} \left[
\frac{27}{4} V \nu_3 \left( ( 2 \beta^2 + 9 \nu_2 ) X_{\rm (ini.)} 
- \sqrt{3} ( 2 \beta^2 + 6 - 3 \nu_2 - 6 \nu_3 ) \psi_{\rm (ini.)} \right) 
\right.
\notag\\
&~~~+ 6 \sqrt{3} \beta^2 \nu_3 ( 2 \nu_1 + \nu_2 ) \dot{X}_{\rm (ini.)} 
+ 18 \beta^2 \nu_2 \nu_3 \dot{\psi}_{\rm (ini.)}
\notag\\
&~~~
- \frac{16}{5} \nu_3 \varepsilon 
\left( \left[ 2 \beta^4 - ( 2 - 9 \nu_2 ) \beta^2 + 27 \nu_2 \right] 
(A_{12} - A_{31}) 
\right.
\notag\\
&~~~
\left.
\left.
+ \sqrt{3} \left[ 2 \beta^4 + ( 4 - 3 \nu_2 - 6 \nu_3 ) \beta^2 
+ 9 ( 2 \nu_1 + \nu_2 ) \right] (B_{12} - B_{31}) \right)
\right] \sin (\beta \bar{t}) ,
\end{align}
with 
$\alpha \equiv \sqrt{(1 + \sqrt{1 - 27 V})/2}$ and 
$\beta \equiv \sqrt{(1 - \sqrt{1 - 27 V})/2}$.

Equations \eqref{ex-X} and \eqref{ex-psi} mean that 
the perturbations $X$ and $\psi$ do not increase with time 
but oscillate around some values. 
As mentioned already, 
the triangular configuration will 
adiabatically shrink and be kept 
in equilibrium.

On the other hand, 
$\chi_{12}$ corresponding to the scale transformation of the system includes 
a linear term in time as 
\begin{align}
\chi_{12}^{(t)} \equiv 
\frac{16}{5} \varepsilon 
\left[
- 2 V B_{12} 
+ \sqrt{3} \nu_2 \nu_3 (A_{12} - A_{31}) 
+ \nu_3 ( 2 - \nu_2 - 2 \nu_3 ) (B_{12} - B_{31})
\right] .
\end{align}
Hence, 
the triangle changes with time as a similarity transformation.
From Eqs. \eqref{ex-AIJ} and \eqref{ex-BIJ}, 
we obtain 
\begin{align}
\chi_{12}^{(t)} &= 
- \frac{64}{5 V} \varepsilon 
\left( \nu_1^2 \nu_2^2 + \nu_2^2 \nu_3^2 + \nu_3^2 \nu_1^2 
- \nu_1^2 \nu_2 \nu_3 - \nu_1 \nu_2^2 \nu_3 - \nu_1 \nu_2 \nu_3^2 
\right) 
\notag\\
&=
- \frac{32}{5 V} \varepsilon 
\left[ \nu_1^2 ( \nu_2 - \nu_3 )^2 
+ \nu_2^2 ( \nu_3 - \nu_1 )^2 + \nu_3^2 ( \nu_1 - \nu_2 )^2 \right] 
\notag\\
&\leq 0 ,
\end{align}
where the equality holds if and only if $\nu_1 = \nu_2 = \nu_3 = 1/3$.
In this equality case, the size of triangle does not change. 
This is because 
gravitational waves are not emitted from the triangular configuration 
as a consequence of a complete phase cancellation of the waves 
in the quadrupole approximation 
when $\nu_1 = \nu_2 = \nu_3 = 1/3$
[see Eqs. \eqref{ex-AIJ} and \eqref{ex-BIJ}]. 
For the Newtonian stable case as in Eq. \eqref{cond.Newton}, 
one can obtain $\chi_{12}^{(t)} < 0$.
Moreover, 
the perturbation $\varpi$ in the orbital frequency also has 
a linear term in time as 
\begin{align}
\varpi^{(t)} &\equiv - \frac{24}{5 V} \varepsilon \left[
- 2 V B_{12} + \sqrt{3} \nu_2 \nu_3 (A_{12} - A_{31}) 
+ \nu_3 ( 2 - \nu_2 - 2 \nu_3 ) (B_{12} - B_{31}) 
\right]
\notag\\
&= - \frac32 \chi_{12}^{(t)}
\notag\\
&\geq 0 .
\end{align}
Substituting this into the first term of Eq. \eqref{ex-varpi}, 
one can see that 
the system shrinks with increasing the orbital frequency linearly in 
the normalized time $\bar{t} \equiv \omega t$.

Before closing this section, 
let us discuss the effects of the 1PN corrections on the long-term stability.
In the long time evolution, 
the 1PN corrections to this triple system will not be negligible.
In fact, 
it has been implied that for some mass ratios, 
even if the Newtonian is stable, 
the triangular configuration in the restricted three-body problem 
may break up as its final fate \cite{SM}.
Hence, 
it is worthwhile to study the long-term stability for three finite masses. 
After a long time (i.e. $\bar{t} \gg 1$), 
the perturbation in the orbital frequency $\varpi$ increases, 
where the linear term in time $\varpi^{(t)}$ dominates 
and the others become negligible. 
Therefore, the orbital frequency can be rewritten as 
\begin{align}
\omega \simeq \omega_{\rm (ini.)} ( 1 + \varpi^{(t)} \bar{t} ) .
\end{align}

\begin{figure}[h]
\begin{center}
  \includegraphics[width=10cm]{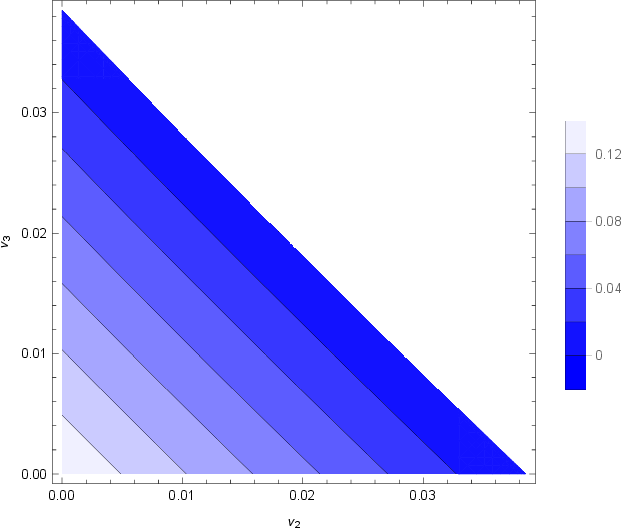}
\caption{
A contour map of the critical values of $\lambda$, 
which are marginal points of Eq. \eqref{CoS1PN}, 
as a function of $\nu_2$ and $\nu_3$.
Note that 
since Eq. \eqref{CoS1PN} is valid only for small values of $\lambda$, 
the lower-left region may not be accurate. 
Indeed, 
one can see $V < \lambda$ in this region; 
thus, the critical values of $\lambda$ are 
very sensitive to higher PN corrections.}
\label{fig-2}
\end{center}
\end{figure}

Figure \ref{fig-2} shows a contour map of the critical values of $\lambda$, 
which are marginal points of Eq. \eqref{CoS1PN}, 
as a function of $\nu_2$ and $\nu_3$.
Note that 
since Eq. \eqref{CoS1PN} is valid only for small values of $\lambda$, 
the lower-left region of Fig. \ref{fig-2} may not be accurate. 
Indeed, 
one can see $V < \lambda$ in this region; 
thus, the critical values of $\lambda$ are 
very sensitive to higher PN corrections.

We also perform numerical tests with an adiabatic treatment 
for two cases of initial values.
Case 1:
$\nu_1 = 0.973, ~~
\nu_2 = 0.027, ~~
\nu_3 = 0 , ~~
\lambda_{\rm (ini.)} = 1/200$, 
which use the same values 
in Fig. 3 of Ref. \cite{SM}.
Case 2:
$\nu_1 = 0.98, ~~
\nu_2 = 0.01 , ~~
\nu_3 = 0.01 , ~~
\lambda_{\rm (ini.)} = 1/200$.
In case 1, 
the system becomes unstable with 
$\bar{t} \sim 10^{5}$ and $\lambda_{\rm (fin.)} \approx 1/25$.
This is consistent with the result in Ref. \cite{SM}.
In case 2, 
the system becomes unstable with 
$\bar{t} \sim 10^{6}$ and $\lambda_{\rm (fin.)} \approx 1/15$.
These results are in agreement with Fig. \ref{fig-2}.
In both cases, 
the systems, which are initially stable, 
become unstable in the final states. 
Therefore, 
it is unlikely that 
the triangular configuration shrinks to merge. 
However, 
in such final states 
where $\lambda_{\rm (fin.)} \sim 0.1$, 
it is necessary to incorporate the higher order PN corrections.
Moreover, it has been pointed out that even for binary systems 
the PN approximation may be no longer valid in such a region \cite{YB,SFN}.
Therefore, 
we need another approach, 
which is valid in strong fields, 
in order to investigate the stability of the system more precisely.

\section{Conclusion}
\label{discussion}

We have investigated the gravitational radiation reaction to 
Lagrange's equilateral triangular orbit.
It has been found that 
the triangular configuration is adiabatically shrinking and 
kept in equilibrium 
with increasing the orbital frequency
at the 2.5PN order 
if the mass ratios satisfy the Newtonian stability condition 
as in Eq. \eqref{CoSN}. 
These results support the assumption in Ref. \cite{Asada}, 
where Lagrange's orbit shrinks and remains an equilateral triangle.
Therefore, 
it may be possible to distinguish a binary source
from a three-body system in Lagrange's orbit or others 
by using Asada's method as a binary source test.

We have also discussed long-term stability involving the 1PN corrections
and shown that 
the triangular configuration, 
which is initially stable, 
will become unstable in the final states where $\lambda_{\rm (fin.)} \sim 0.1$.
It is left as a future work 
to investigate the dynamics of the three bodies more precisely.

\section*{Acknowledgments}

We would like to thank Kota Iseki and Naoya Harada for 
discussions of the early stage of this work. 
We are grateful to Yuuiti Sendouda, Takahiro Tanaka, Hiroyuki Nakano, 
and Naoki Seto for useful comments.
This work was supported in part by JSPS Grant-in-Aid for JSPS Fellows, 
No. 15J01732 (K.Y.) and 
JSPS Grant-in-Aid for Scientific Research, No. 26400262 and No. 15H00772 (H.A.).

\appendix
\section{A DERIVATION OF 
GRAVITATIONAL RADIATION REACTION FORCE}
\label{app1}

The 2.5PN correction to the metric $h^{Q}_{0 0}$ in the harmonic gauge is 
\cite{MTW, Maggiore, Blanchet} 
\begin{align}
h^{Q}_{0 0} = - \frac{2 \Phi}{c^2} ,
\end{align}
where 
\begin{align}
\Phi \equiv \frac{G}{5 c^5} \frac{d^5 Q_{i j}}{d t^5} x^i x^j 
\end{align}
is the correction to the Newtonian potential with the mass quadrupole moment
\begin{align}
Q_{i j} \equiv 
\int \rho \left( x_i x_j - \frac13 \delta_{i j} r^2 \right) d^3 x .
\end{align}
Thus, the quadrupole radiation reaction force per unit mass is 
\begin{align}
F_i^{\rm RR} = - \frac{\partial \Phi}{\partial x^i} .
\label{FRR-origin}
\end{align}

We consider Lagrange's orbit of the three bodies on the $(x,y)$ plane, 
where nonzero components of the quadrupole moment $Q_{i j}$ are
\begin{align}
Q_{x x} &= \sum_J m_J \left( r_J \right)^2 \cos^2 \theta_J 
+ {\rm constant} , \\
Q_{x y} &= Q_{y x} = 
\sum_J m_J \left( r_J \right)^2 \cos \theta_J \sin \theta_J , \\
Q_{y y} &= \sum_J m_J \left( r_J \right)^2 \sin^2 \theta_J 
+ {\rm constant} , \\
Q_{z z} &= {\rm constant} .
\end{align}
Therefore, 
one can see 
\begin{align}
F_z^{\rm RR} \propto 
\frac{d^5 Q_{z k}}{d t^5} x^k 
= \frac{d^5 Q_{z z}}{d t^5} x^z = 0 .
\end{align}
It follows that 
the orbital plane is not changed by the radiation reaction, 
and hence, 
we focus on the $(x,y)$ plane in the following.

The reaction force \eqref{FRR-origin} 
on a field point $\bs{r} = r (\cos \theta, \sin \theta)$ 
can be expressed as 
\begin{align}
\bs{F}^{\rm RR} &=
\frac{32}{5} \frac{G r \omega^5}{c^5} \sum_J m_J r_J^2
\begin{bmatrix}
\sin (2 \theta_J) & - \cos (2 \theta_J) \\
- \cos (2 \theta_J) & - \sin (2 \theta_J)
\end{bmatrix}
\begin{bmatrix}
\cos \theta \\
\sin \theta
\end{bmatrix} 
\notag\\
&=
\frac{32}{5} \frac{G r \omega^5}{c^5} \sum_J m_J r_J^2
\begin{bmatrix}
\sin (2 \theta_J - 2 \theta) & \cos (2 \theta_J - 2 \theta) \\
- \cos (2 \theta_J - 2 \theta) & \sin (2 \theta_J - 2 \theta)
\end{bmatrix}
\begin{bmatrix}
\cos \theta \\
\sin \theta
\end{bmatrix} .
\end{align}
By replacing $r$ and $\theta$ with $r_I$ and $\theta_I$, 
respectively, 
the force to the $I$th body per unit mass is 
\begin{align}
F^{\rm RR}_I = - \frac{32}{5} \frac{G M}{\ell^2} \varepsilon \bar{r}_I  
[ A_I + i B_I ] e^{i \theta_I} ,
\end{align}
where we define 
\begin{align}
\varepsilon \equiv 
\left( \frac{G M \omega}{c^3} \right)^{5/3} ,
\end{align}
and 
\begin{align}
A_I &\equiv \sum_J \nu_J ( \bar{r}_J )^2 
\sin (2 \theta_I - 2 \theta_J) , \\
B_I &\equiv \sum_J \nu_J ( \bar{r}_J )^2 
\cos (2 \theta_I - 2 \theta_J) .
\end{align}

\section{SOLVING 
THE EQUATIONS OF MOTION FOR PERTURBATIONS}
\label{solve}

In order to solve Eqs. \eqref{EoM12Re}--\eqref{EoM31Im}, 
let us take the Laplace transform as
\begin{align}
F (s) \equiv \mathscr{L} [f (t)] = \int_0^{\infty} e^{- s t} f (t) d t .
\end{align}
Thus, 
the perturbed equations of motion \eqref{EoM12Re}--\eqref{EoM31Im} become
\begin{align}
(s^2 - 3) L_{12} - s \chi_{12 \, {\rm (ini.)}} - \dot{\chi}_{12 \, {\rm (ini.)}} 
- 2 L_{\varpi} - \frac94 \nu_3 L_X - \frac{3 \sqrt{3}}{4} \nu_3 L_{\psi} 
- \frac{16}{5} \varepsilon A_{12} \frac{1}{s} &= 0 , 
\label{eq1} \\
2 s L_{12} - 2 \chi_{12 \, {\rm (ini.)}} + s L_{\varpi} - \varpi_{\rm (ini.)}  
- \frac{3 \sqrt{3}}{4} \nu_3 L_X + \frac94 \nu_3 L_{\psi} 
+ \frac{16}{5} \varepsilon B_{12} \frac{1}{s} &= 0 , 
\label{eq2} \\
(s^2 - 3) L_{12} - s \chi_{12 \, {\rm (ini.)}} - \dot{\chi}_{12 \, {\rm (ini.)}} 
- 2 L_{\varpi} + \left( s^2 - 3 + \frac94 \nu_2 \right)  L_X 
\notag\\
- s X_{\rm (ini.)} 
- \dot{X}_{\rm (ini.)} - \left( 2 s + \frac{3 \sqrt{3}}{4} \nu_2 \right) L_{\psi} 
+ 2 \psi_{\rm (ini.)} - \frac{16}{5} \varepsilon A_{31} \frac{1}{s} &= 0 , 
\label{eq3} \\
2 s L_{12} - 2 \chi_{12 \, {\rm (ini.)}} + s L_{\varpi} - \varpi_{\rm (ini.)} 
+ \left( 2 s - \frac{3 \sqrt{3}}{4} \nu_2 \right) L_X 
\notag\\
- 2 X_{\rm (ini.)} 
+ \left( s^2 - \frac94 \nu_2 \right) L_{\psi} - s \psi_{\rm (ini.)} 
- \dot{\psi}_{\rm (ini.)} + \frac{16}{5} \varepsilon B_{31} \frac{1}{s} &= 0 , 
\label{eq4}
\end{align}
where the subscript (ini.) means the initial value 
and we define 
\begin{align}
L_{12} (s) &\equiv \mathscr{L} [\chi_{12} (t)] , \\
L_{\varpi} (s)  &\equiv \mathscr{L} [\varpi (t)] , \\
L_X (s) &\equiv \mathscr{L} [X (t)] , \\
L_{\psi} (s)  &\equiv \mathscr{L} [\psi (t)] , 
\end{align}
for simplicity.

Subtracting Eqs. \eqref{eq1} and \eqref{eq2} from 
Eqs. \eqref{eq3} and \eqref{eq4}, respectively, 
we obtain 
\begin{align}
\left( s^2 - 3 + \frac94 ( \nu_2 + \nu_3 ) \right)  L_X 
- \left( 2 s + \frac{3 \sqrt{3}}{4} (\nu_2 - \nu_3) \right) L_{\psi} 
\notag\\
- X_{\rm (ini.)} s - \dot{X}_{\rm (ini.)} + 2 \psi_{\rm (ini.)} 
+ \frac1s \frac{16}{5} \varepsilon (A_{12} - A_{31}) 
&= 0 , \\
\left( 2 s - \frac{3 \sqrt{3}}{4} (\nu_2 - \nu_3) \right) L_X 
+ \left( s^2 - \frac94 (\nu_2 + \nu_3) \right) L_{\psi} 
\notag\\
- \psi_{\rm (ini.)} s - ( 2 X_{\rm (ini.)} + \dot{\psi}_{\rm (ini.)} ) 
- \frac1s \frac{16}{5} \varepsilon (B_{12} - B_{31}) 
&= 0 .
\end{align}
These can be solved for $L_X$ and $L_{\psi}$ as 
\begin{align}
L_X &= \frac{X_{\rm (ini.)} s^4 + \dot{X}_{\rm (ini.)} s^3 + g_2 s^2 + g_1 s + g_0}
{s (s - i \alpha) (s + i \alpha) (s - i \beta) (s + i \beta)} , \\
L_{\psi} &= \frac{\psi_{\rm (ini.)} s^4 + \dot{\psi}_{\rm (ini.)} s^3 
+ h_2 s^2 + h_1 s + h_0}
{s (s - i \alpha) (s + i \alpha) (s - i \beta) (s + i \beta)} ,
\end{align}
where 
\begin{align}
g_2 &= \left( 4 - \frac94 (\nu_2 + \nu_3) \right) X_{\rm (ini.)} 
+ \frac{3 \sqrt{3}}{4} ( \nu_2 - \nu_3 ) \psi_{\rm (ini.)} 
+ 2 \dot{\psi}_{\rm (ini.)} 
- \frac{16}{5} \varepsilon (A_{12} - A_{31}) , \\
g_1 &= \frac{3 \sqrt{3}}{2} (\nu_2 - \nu_3) X_{\rm (ini.)} 
- \frac94 ( \nu_2 + \nu_3 ) \dot{X}_{\rm (ini.)} 
\notag\\
&~~~
+ \frac92 (\nu_2 + \nu_3) \psi_{\rm (ini.)} 
+ \frac{3 \sqrt{3}}{4} ( \nu_2 - \nu_3 ) \dot{\psi}_{\rm (ini.)} 
+ \frac{32}{5} \varepsilon (B_{12} - B_{31}) , \\
g_0 &= \frac{12}{5} \varepsilon 
\left( 
3 (\nu_2 + \nu_3) (A_{12} - A_{31}) + \sqrt{3} (\nu_2 - \nu_3) (B_{12} - B_{31})
\right) , \\
h_2 &= \frac{3 \sqrt{3}}{4} (\nu_2 - \nu_3) X_{\rm (ini.)} 
- 2 \dot{X}_{\rm (ini.)} 
+ \left( 1 + \frac94 (\nu_2 + \nu_3) \right) \psi_{\rm (ini.)} 
+ \frac{16}{5} \varepsilon (B_{12} - B_{31}) , \\
h_1 &= - \left( 6 - \frac92 (\nu_2 + \nu_3) \right) X_{\rm (ini.)} 
+ \frac{3 \sqrt{3}}{4} (\nu_2 - \nu_3) \dot{X}_{\rm (ini.)} 
\notag\\
&~~~
- \frac{3 \sqrt{3}}{2} (\nu_2 - \nu_3) \psi_{\rm (ini.)} 
- \left( 3 - \frac94 (\nu_2 + \nu_3) \right) \dot{\psi}_{\rm (ini.)} 
+ \frac{32}{5} \varepsilon (A_{12} - A_{31}) , \\
h_0 &= - \frac{12}{5} \varepsilon 
\left( 
\sqrt{3} (\nu_2 - \nu_3) (A_{12} - A_{31}) 
+ ( 4 - 3 \nu_2 - 3 \nu_3) (B_{12} - B_{31})
\right) .
\end{align}
Moreover, by using these expressions, 
$L_{12}$ and $L_{\varpi}$ are 
\begin{align}
L_{12} &= 
\frac{1}{s^2 (s^2 + 1)} 
\left[
\chi_{12 \, {\rm (ini.)}} s^3 + \dot{\chi}_{12 \, {\rm (ini.)}} s^2 
+ \left(4 \chi_{12 \, {\rm (ini.)}} + 2 \varpi_{\rm (ini.)} 
+ \frac{16}{5} \varepsilon A_{12} \right) s 
- \frac{32}{5} \varepsilon B_{12} \right]
\notag\\
&~~~
+ \frac{3 \left( 3 s + 2 \sqrt{3} \right) \nu_3}{4 s (s^2 + 1)} L_X
+ \frac{3 \left( \sqrt{3} s - 6 \right) \nu_3}{4 s (s^2 + 1)} L_{\psi} , \\
L_{\varpi} &= 
\frac{1}{s^2 (s^2 + 1)} 
\left[
\varpi_{\rm (ini.)} s^3 
- \left( 2 \dot{\chi}_{12 \, {\rm (ini.)}} 
+ \frac{16}{5} \varepsilon B_{12} \right) s^2
- \left( 6 \chi_{12 \, {\rm (ini.)}} + 3 \varpi_{\rm (ini.)} 
+ \frac{32}{5} \varepsilon A_{12} \right) s 
+ \frac{48}{5} \varepsilon B_{12} \right]
\notag\\
&~~~
+ \frac{3 \sqrt{3} \left( s^2 - 2 \sqrt{3} s - 3 \right) \nu_3}
{4 s^2 (s^2 + 1)} L_X 
- \frac{3 \left(3 s^2 + 2 \sqrt{3} s - 9 \right) \nu_3}{4 s (s^2 + 1)} 
L_{\psi} . 
\end{align}
Finally, 
taking the inverse Laplace transform,
one can obtain the solutions Eqs. \eqref{ex-X}--\eqref{ex-varpi}.

\begin{thebibliography}{99}

\bibitem{GWPRL}
B. P. Abbott {\it et al.} 
(LIGO Scientific Collaboration and Virgo Collaboration), 
Phys. Rev. Lett. {\bf 116}, 061102 (2016).
\bibitem{aVIRGO}
F. Acernese {\it et al.} (VIRGO Collaboration), 
Classical Quantum Gravity {\bf 32}, 024001 (2015).
\bibitem{KAGRA}
http://gwcenter.icrr.u-tokyo.ac.jp/researcher/parameters
\bibitem{aLIGO}
J. Aasi {\it et al.} (LIGO Scientific Collaboration), 
Classical Quantum Gravity {\bf 32}, 074001 (2015).
\bibitem{NR}
M. Shibata and K. Taniguchi, Living Rev. Relativity {\bf 14}, 6 (2011).
\bibitem{PN}
T. Futamase and Y. Itoh, Living Rev. Relativity {\bf 10}, 2 (2007). 
L. Blanchet, Living Rev. Relativity {\bf 17}, 2 (2014); 
\bibitem{ST}
M. Sasaki and H. Tagoshi, Living Rev. Relativity {\bf 6}, 6 (2003). 
\bibitem{Damour}
T. Damour and A. Nagar, Phys. Rev. D {\bf 77}, 024043 (2008); 
T. Damour, A. Nagar, E. N. Dorband, D. Pollney, and L. Rezzolla, 
Phys. Rev. D {\bf 77}, 084017 (2008).
\bibitem{CIA}
T. Chiba, T. Imai, and H. Asada, 
Mon. Not. R. Astron. Soc. {\bf 377}, 269 (2007);
\bibitem{GB}
P. Galaviz and B. Br\"{u}gmann, 
Phys. Rev. D {\bf 83}, 084013 (2011). 
\bibitem{Seto1}
N. Seto, Phys. Rev. D {\bf 85}, 064037 (2012).
\bibitem{DSH}
V. Dmitra\v{s}inovi\'{c}, M. \v{S}uvakov, and A. Hudomal, 
Phys. Rev. Lett. {\bf 113}, 101102 (2014).
\bibitem{THA}
Y. Torigoe, K. Hattori, and H. Asada, 
Phys.\ Rev.\ Lett.\  {\bf 102}, 251101 (2009).
\bibitem{Asada}
H. Asada, Phys. Rev. D {\bf 80} 064021 (2009). 
\bibitem{SM}
N. Seto and T. Muto, Phys. Rev. D {\bf 81} 103004 (2010).
\bibitem{Schnittman}
J. D. Schnittman, Astrophys. J. {\bf 724} 39 (2010). 
\bibitem{IYA}
T. Ichita, K. Yamada, and H. Asada, Phys. Rev. D {\bf 83}, 084026 (2011).
\bibitem{YA3}
K. Yamada and H. Asada, Phys. Rev. D {\bf 86}, 124029 (2012).
\bibitem{YTA}
K. Yamada, T. Tsuchiya, and H. Asada, 
Phys. Rev. D {\bf 91}, 124016 (2015).
\bibitem{BDEDSG}
E. Battista {\it et al.}, Phys. Rev. D {\bf 92}, 064045 (2015).
\bibitem{Ransom}
S. M. Ransom {\it et al.}, Nature {\bf 505}, 520 (2014).
\bibitem{BLS}
O. Blaes, M. H. Lee, and A. Socrates, Astrophys. J. {\bf 578}, 775 (2002).
\bibitem{MH}
M. C. Miller and D. P. Hamilton, Astrophys. J. {\bf 576}, 894 (2002).
\bibitem{Wen}
L. Wen, Astrophys. J. {\bf 598}, 419 (2003)
\bibitem{Thompson}
T. A. Thompson, Astrophys. J. {\bf 741}, 82 (2011).
\bibitem{Seto2}
N. Seto, Phys. Rev. Lett. {\bf 111}, 061106 (2013).
\bibitem{Gascheau}
G. Gascheau, C. R. Acad. Sci. {\bf 16}, 393 (1843).
\bibitem{YB}
N. Yunes and E. Berti, Phys. Rev. D {\bf 77}, 124006 (2008).
\bibitem{SFN}
N. Sago, R. Fujita, and H. Nakano have investigated 
accuracy of the PN approximation for extreme mass ratio inspirals
(arXiv:1601.02174[gr-qc]).
They have found that 
there are several local maximums of the relative error
in regions of validity for relatively low-PN order results, 
while regions of validity become larger for higher PN order results.
Hence, higher PN order calculations are needed 
for three-body systems as well as binaries.
\bibitem{MTW}
C. W. Misner, K. S. Thorne, and J. A. Wheeler, 
{\it Gravitation}, 
(Freeman, New York, 1973).
\bibitem{Maggiore}
M. Maggiore, 
{\it Gravitational Waves}
(Oxford University, New York, 2008).
\bibitem{Blanchet}
L. Blanchet, Phys. Rev. D {\bf 55}, 714 (1997).
\end{thebibliography}
\end{document}